\def\ispreprint{1}  % turn to zero if you want for submission to Icarus, turn to 1 if you want a preprint style

\if \ispreprint1
\documentclass[preprint,authoryear,5p]{elsarticle}  % preprint style
\else
\documentclass[preprint,authoryear,12pt]{elsarticle}  %  for submission
\fi

\usepackage[utf8]{inputenc}
\usepackage{lineno} % needed for submission to Icarus
\usepackage{amssymb}
\usepackage{amsmath}
\usepackage{adjustbox}
\usepackage{graphicx,subcaption} 
\usepackage{hyperref} 
\usepackage{setspace}

\begin{document}

\title{Rearrangement of Granular Surfaces on Asteroids due to Thermal Cycling}
\author[a1]{Danielle Bovie}
\ead{dbovie@ur.rochester.edu}
\author[a2]{A. C. Quillen}
\ead{alice.quillen@rochester.edu}
\author[a3]{Rachel Glade}
\ead{rglade@ur.rochester.edu}

\address[a1]{Material Science Program, University of Rochester, Rochester, NY 14627, USA}

\address[a2]{Department of Physics and Astronomy, University of Rochester, Rochester, NY 14627, USA}

\address[a3]{Department of Earth and Environmental Sciences, University of Rochester, Rochester, NY 14627, USA}

\begin{abstract}
In granular systems, thermal cycling causes compaction, creep, penetration of dense objects, and ratcheting of grains against each other. On asteroid surfaces, thermal cycling is high amplitude and can happen billions of times in a few million years. We use a 1-dimensional thermophysical conductivity model to estimate the relative displacement of grains in proximity to one another, caused by variations in thermal conductivity or shadows. We find that grains would experience relative displacements of order a few microns during each thermal cycle. If thermal cycling causes diffusive transport, then the asteroid's few centimeters deep thermal skin depth could flow a few centimeters in a million years.  Thermal cycling could cause long-distance flows on sloped surfaces, allowing fine materials to collect in depressions. 

\end{abstract}

\maketitle

\begin{abstract}
    
\end{abstract}

%\section{Directories}

%Oblique Mid Vel (Bingcheng's directory) \url{https://drive.google.com/drive/folders/0ACr4yphwu_tQUk9PVA}

%Manual for Chronos 2.1 \url{https://docslib.org/doc/11428205/chronos-2-1-hd-user-manual}

%Data of 1000fps movies and laser line scans (Alice's directory) \url{https://drive.google.com/drive/folders/1ARi-WdDD3QD56vgFhx3V_301NIWY-cjV?usp=sharing}

%Data for scanned lasers is actually on hard drives because it would not fit on the google drives. 

\section{Introduction}

\label{chap:introduction}
Most asteroids rotate rapidly, with spin rotational periods of only a few hours \citep{Pravec_2008}. Asteroids have little atmosphere to insulate their surfaces, hence asteroids experience sizable hundred Kelvin diurnal variations in surface temperature. Asteroid surfaces host heterogeneous granular systems (e.g., \cite{Walsh_2018, DellaGiustina_2019, Sugita_2019}), thus understanding the thermal response of granular systems is key to understanding the evolution of their surfaces.

Thermophysical modeling of optical and infrared light curves gives information about the composition and structure of asteroid surfaces through measurements of thermal inertia;
\begin{equation}
\Gamma = \sqrt{k_T \rho C_P}
\end{equation}
where $k_T$ is the thermal conductivity, $\rho$ is the mean density, and $C_P$ is the specific heat. Thermal inertia is a material's ability to resist temperature change \citep{Rozitis_2020}. Measured values of an asteroid’s thermal inertia can span a large range from about 10 to 1000 J m$^{-2}$ s$^{-1/2}$ K$^{-1}$ \citep{Delbo_2009, Rozitis_2011,Delbo_2015, Harris_2020}. A surface comprised of a small-grained granular material is predicted to have lower thermal inertia than solid rock (e.g., \cite{Gundlach_2013}). Surfaces covered in fine regolith have extremely low measured thermal inertia values \citep{Delbo_2009, Rozitis_2011,Gundlach_2013, Murdoch_2015,Harris_2020}. For example, asteroid (21) Lutetia has $\Gamma\le 30$ J m$^{-2}$ s$^{-1/2}$ K$^{-1}$ \citep{Keihm_2012} which is attributed to a thick layer of fine regolith \citep{Gundlach_2013}. Porosity can also affect thermal inertia. From imaging of the asteroid (162173) Ryugu, some apparently solid boulders also have low thermal inertia, implying that these boulders have low porosity \citep{Grott_2019}.

The surfaces of asteroid (101955) Bennu and asteroid (162173) Ryugu exhibit a large scatter in their local thermal inertia values, values measured at different positions on the asteroid surface, ranging from about 200 to 400 J m$^{-2}$ s$^{-1/2}$ K$^{-1}$  \citep{Rozitis_2020, Shimaki_2020}. The heterogeneity of the surface and this wide range in local thermal inertia values implies that materials with different thermal properties are often in proximity. When materials with different thermal expansion coefficients are in contact, thermal stress due to thermal cycling can cause cumulative damage when microscopic cracks widen and are extended. This phenomenon is known as "thermal fatigue" \citep{Delbo_2014} and is a fragmentation and regolith generation mechanism. While previous studies have focused on fracturing within solids on the surfaces of asteroids, here we focus on the connection between thermal cycling and the potential for dynamics within the granular system. When a granular medium with grains of various sizes is subjected to vibrations grains will experience size segregation. Larger objects can rise to the top regardless of density \citep{rosato1987}. This is called the Brazil Nut Effect and has been theorized as a mechanism for grain sorting on asteroid surfaces \citep{Hestroffer_2019} caused by vibrations from impacts and seismic activity \citep{Asphaug2001}. While thermal cycling has not been found to induce the Brazil Nut Effect, the expansion and contraction of grains due to thermal cycling might induce granular flow.

The effect of temperature variations on a granular medium has been examined and discussed in laboratory settings for industrial applications \citep{Divoux_2010, Chen_2009, Blanc_2013, Percier_2013}. Thermal cycling in granular systems causes a variety of phenomena, including rearrangements of the force chain network \citep{Clement_1997, Wang_2018},  compaction \citep{Chen_2006, Vargas_2007, Chen_2009},  creep \citep{Divoux_2008, Percier_2013, Blanc_2013, Deshpande_2021}, fluctuations in transport properties \citep{Liu_1992}, and penetration of dense objects \citep{Chen_2009}. Asteroid surfaces can experience a large day-night variation in temperature, $\Delta T$, and while strains caused by a single thermal cycle are small, each thermal cycle occurs once per rotation period.  As rotation periods are hours to days \citep{Pravec_2008, Harris_2009}, during the 10 million year lifetime of a near-earth asteroid \citep{Gladman_2000}, a billion cycles could occur.  Effects associated with thermal cycling could be significant if they accumulate over time. 

In this manuscript, we explore how thermal cycling might affect the heterogeneous granular systems that exist on asteroid surfaces by using 1-dimensional thermophysical models. We explore the role of inhomogeneous thermal conductivity and shadows in affecting variations of temperature as a function of depth and the associated relative thermal expansion. Using this we estimate the relative motions of the material as it undergoes thermal cycling.  Relative displacements from the thermophysical models are used to estimate how far grains could move with respect to one another. If grains are able to ratchet against one another or this relative motion is occurring on a slope, permanent relative motions would accumulate over the many thermal cycles that an asteroid experiences.

\section{Relative Displacements due to Heterogeneous Thermal Properties}
\label{chap:Displacements}
We adopt a 1-dimensional thermal conductivity model, following the procedure that is summarized in section 2.2 of \citet{Rozitis_2011}. The model integrates the heat diffusion equation vertically (in $z$) with an energy-balanced surface boundary condition that takes into account the absorption and radiation of solar energy. We apply a zero heat flux lower boundary condition. The resulting model gives temperature $T(z,\tau)$ where $z>0$ is depth below the surface and $\tau \in [0,1]$ specifies the phase of the rotation of the asteroid. Additional details of this model are described in \ref{sec:append}.

\subsection{Fiducial model}
We adopt a fiducial thermophysical model and then consider variations from it. The parameters for the fiducial model are listed in Table \ref{tab:fiducial}. The parameters used for the fiducial model follow the review of thermophysical properties given by \citet{Harris_2002} and \citet{Delbo_2015}. We used a typical value for specific heat intensity $C_P \sim 500\ {\rm J\ kg}^{-1} K^{-1}$ based on measurements of stony meteorites \citep{Consolmagno_2013}. For density, we adopt a typical value of $\rho = 2000\ {\rm kg\ m}^{-3}$ based on the compilation by \citet{Carry_2012} and recent asteroid thermophysical models \citep{Hung_2022}. For the rotational period, we select $P = 6$ hours, within the range measured from surveys of asteroid light curves \citep{Pravec_2008}. For the bond albedo, we use $A_B = 0.1$, typical of asteroid thermophysical models \citep{Hung_2022}. For thermal emissivity, we use $\epsilon = 1$. For the linear thermal expansion coefficient we adopt $\alpha_L = 10^{-5} K^{-1}$ following measurements for common minerals by \citet{Hazen_1977}. The fiducial model has a thermal skin depth of 5.2 cm where the thermal skin depth, 
\begin{equation}
    l_{2\pi} \equiv \sqrt{4 \pi P_{rot} \kappa},
\end{equation}
characterizes the depth of the surface layer that undergoes thermal cycling. The thermal diffusivity $\kappa$ depends on the thermal conductivity, density, and specific heat;
\begin{equation}
    \kappa \equiv \frac{k_T}{\rho C_P}. \label{eqn:kappa}
\end{equation}
Because the variations in mean density and specific heat are not expected to be large, the large range in measurements of thermal inertia on asteroid surfaces is interpreted to arise from variations in thermal conductivity that are associated with layers of fine dust \citep{Delbo_2009,Harris_2020,Hung_2022} or variations in rock porosity \citep{Grott_2019,Rozitis_2020}. 

Because we are not varying porosity in our model we often equate thermal inertia and thermal conductivity. Global maps of thermal inertia on asteroid Bennu \citep{Rozitis_2020} reveal that the surface is heterogeneous. The areas with the lowest thermal inertia ($\sim$ 180 to 250 J m$^{-2}$ K$^{-1}$ s$^{-1/2}$) on Bennu are populated by large, low-reflectance grains, whereas areas with higher thermal inertia ($\sim$ 350 to 400 J m$^{-2}$ K$^{-1}$) have mixtures of low and high-reflectance grains. This suggests that grains with very different thermal conductivity can come into contact.  The ratio of lowest to highest thermal inertia on Bennu is about 1.7, corresponding to a ratio of lowest to highest thermal conductivity of about 3. We use this amplitude of variations to estimate the relative displacements of materials with different conductivities.

\begin{table}
    \caption{Fiducial model parameters}
    \begin{adjustbox}{width=\columnwidth,center}
    
    \begin{tabular}{llll}
    \hline
    Quantity & Symbol &  Value \\
    \hline
    Spin rotation period & $P_{rot}$ & 6 hr \\
    Heliocentric distance  & $R_H$  &  1 AU \\
    Bond albedo & $A_B$ & 0.1 \\
    Thermal emissivity & $\varepsilon$ & 1 \\
    Density & $\rho$ & 2000 kg m$^{-3}$ \\
    Specific heat & $C_P$ & 500 J kg$^{-1}$ K$^{-1}$  \\
    Thermal conductivity & $k_T$ & $10^{-2}$  W m$^{-1}$ K$^{-1}$ \\
    Thermal diffusivity & $\kappa=k_T/(\rho C_P)$ &  $10^{-8}$ m$^2$ s$^{-1}$ \\
    Thermal inertia & $\Gamma = \sqrt{\rho C_P k_T}$ & 100  K  m$^{-2}$ s$^{- \frac{1}{2}}$ K$^{-1}$
    \\
    Thermal skin depth & $l_{2\pi} = \sqrt{4 \pi P_{rot} \kappa} $ & 5.2 cm \\
    Linear thermal expansion coefficient & $\alpha_L$ & $10^{-5}$ K$^{-1}$ \\
    Illumination function & $\Psi(t)$ &  Equation \ref{eqn:illum} \\ 
    Shadow function & $S(t)$ & 0 \\
%         &   & \\
    \hline
    \end{tabular}
    \end{adjustbox}
    \label{tab:fiducial}
    %\\
    %{Notes: The illumination function for the fiducial model is given in equation \ref{eqn:illum}}
\end{table}

\subsection{Sensitivity of the thermal model to variations in thermal
conductivity}
\begin{figure}[t]
    \centering
    \includegraphics[height=40mm, trim = 14mm 0 0 0, clip]{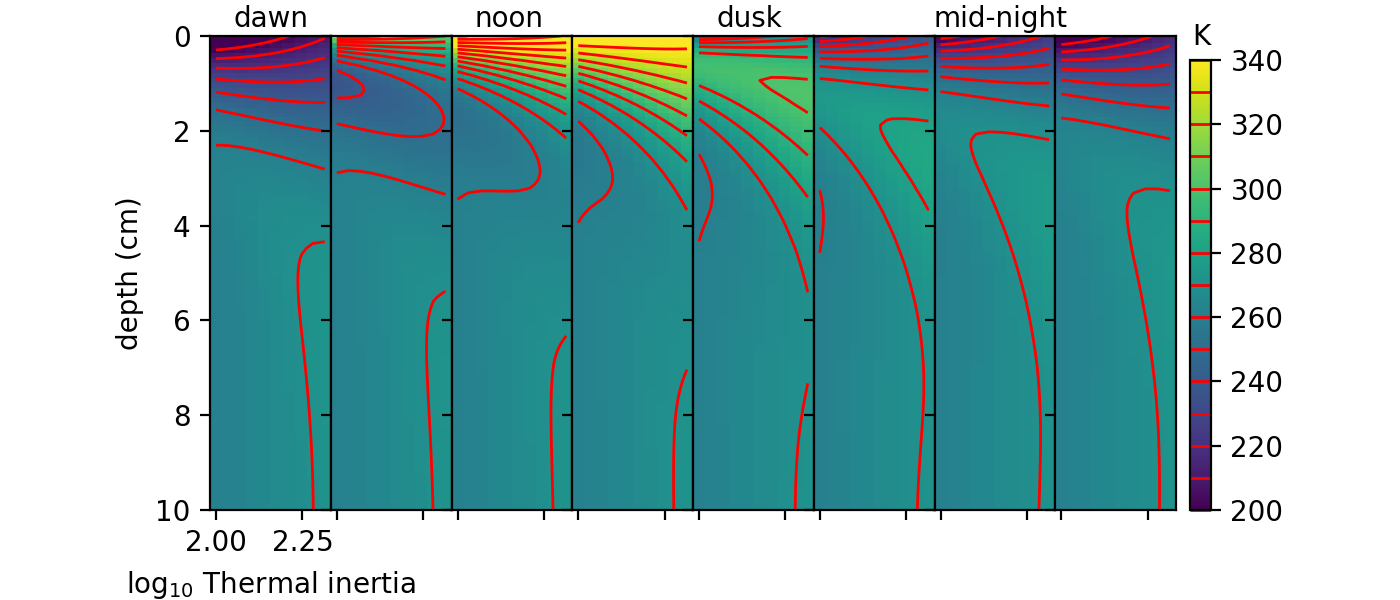}
    \caption{Shown is the variation of temperature with depth below the surface as a function of thermal inertia $\Gamma$. Each panel shows a different phase of rotation. The x-axis shows $\log_{10} \Gamma$ with thermal inertia in units of J m$^{-2}$ s$^{-1/2}$ K$^{-1}$. The y-axis shows depth in units of centimeters. The colorbar shows temperature in Kelvin. Parameters of the model are taken from  Table \ref{tab:fiducial}, except the thermal conductivity, which is varied to give a range in thermal inertia $\Gamma$. Due to differences in thermal conductivity, temperature contours are not horizontal. Columns of material with different conductivity would have different temperature profiles.}
    \label{fig:k_T}
\end{figure}

We vary the thermal conductivity $k_T$ from our thermophysical model to see how temperature as a function of depth and rotational phase is sensitive to  conductivity. For each value of thermal conductivity $k_T$ we compute the temperature as a function of depth $z$ and rotation phase $\tau$. This gives us a three-dimensional array $T(z,k_T,\tau)$. The resulting temperature as a function of depth for eight different orbital phases is shown in Figure \ref{fig:k_T} with physical parameters for the thermophysical model listed in Table \ref{tab:fiducial} but with thermal conductivity $k_T$ ranging from $10^{-2}$ to $5 \times 10^{-2}$ W m$^{-1}$ K$^{-1}$. We select a range for thermal conductivity that matches the range in thermal inertia measured on Bennu's surface. In Figure \ref{fig:k_T}, the x-axes show the log of thermal inertia, the y-axis shows depth, the colorbar shows temperature, and each panel shows a different rotation phase. For the model shown in Figure \ref{fig:k_T}, the skin depth for the lowest thermal conductivity is about 5 cm and is twice as large as for the highest thermal conductivity. Each panel shows the temperature profile as a function of depth and thermal conductivity. Because the skin depth is a function of thermal conductivity, below the surface two substrates with different thermal conductivities would have different temperature profiles. At a depth of about 2 cm, the highest conductivity material is cooler at dawn and hotter past noon, than the lowest conductivity material. These temperature differences would cause variations in thermal expansion. The hotter material expands more than the cooler material and this can cause grains or pebbles in proximity to rub or slide against each other.

\begin{figure}[t]
    \centering
    \includegraphics[height=50mm]{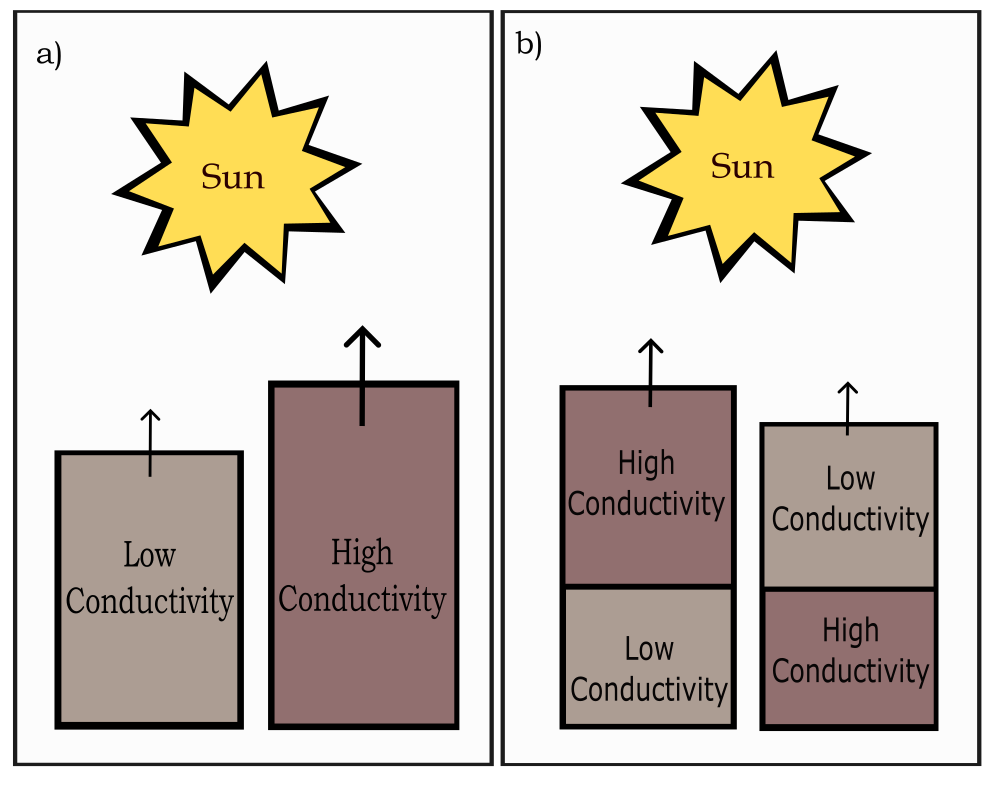}
    \caption{a) Shown is the setup for two materials with different thermal conductivities next to each other and their difference in expansion under direct sunlight. b) Shown is the setup for materials with different conductivities stacked vertically and their difference in expansion under direct sunlight.}
    \label{fig:kt_picture}
\end{figure}

\subsection{Height variations as a function of thermal conductivity}
We use our thermophysical models of temperature as a function of depth $T(z)$ to estimate variations in height caused by thermal expansion. Because temperature variations are small below the thermal skin depth, we assume that the base of each column of material is fixed. We use the differences in column height to characterize the size of the relative motions of grains in proximity that are caused by differences in thermal properties, such as thermal conductivity.
This is illustrated in Figure \ref{fig:kt_picture}, where two columns of material with different conductivity are next to one another and rub against each other due to variations in thermal expansion. 

Vertical displacement in a column of material caused by thermal expansion is estimated by integrating displacement from the base where the temperature is nearly constant; 
\begin{equation}
    \delta_z(z,k_T,\tau) =
    \int_{base}^z dz \ \alpha_L ( T(z,k_T,\tau) - T_{base} ) . \label{eqn:dz}
\end{equation}
This is an estimate for the vertical displacement as a function of depth $z$, at any particular rotational phase $\tau$, and as a function of thermal conductivity $k_T$. % assuming that the base of each vertical column is fixed. 

\begin{figure}[t]
    \centering
    \includegraphics[height=40mm,  trim = 10mm 0 0 0, clip]{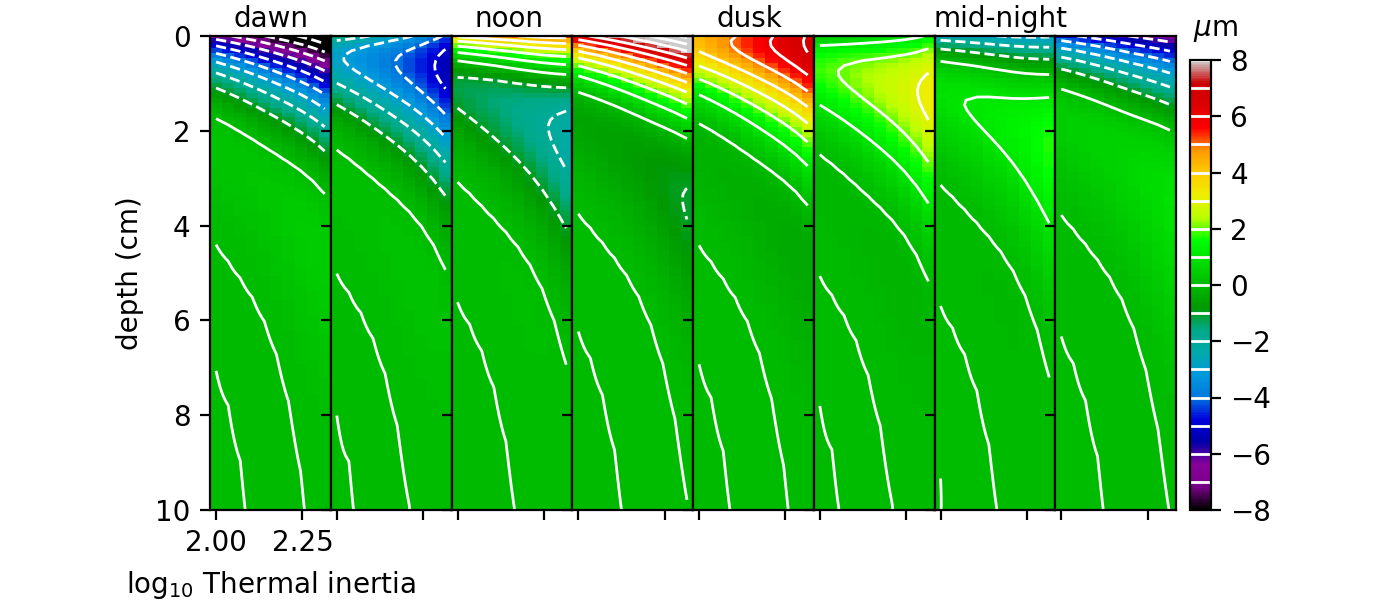}
    \caption{Similarly to Figure \ref{fig:k_T}, depth below the surface is shown as a function of thermal inertia $\Gamma$. However, instead of temperature, the integrated vertical displacement is shown, computed using Equation \ref{eqn:dz}. Each panel shows a different phase of rotation. The x-axis shows $\log_{10} \Gamma$ with thermal inertia in units of J m$^{-2}$ s$^{-1/2}$ K$^{-1}$. The y-axis shows depth in units of centimeters. The colorbar shows the size of the displacement in microns. This figure was computed using a thermal expansion coefficient of $\alpha_L = 10^{-5}$ K$^{-1}$. Contours are not horizontal, indicating that two columns of material with different conductivities in proximity would rub against each other due to relative displacements of order a few microns.}
    \label{fig:dz}
\end{figure}

\begin{figure}[t]
    \centering
    \includegraphics[height=40mm,trim = 10mm 0 0 0, clip]{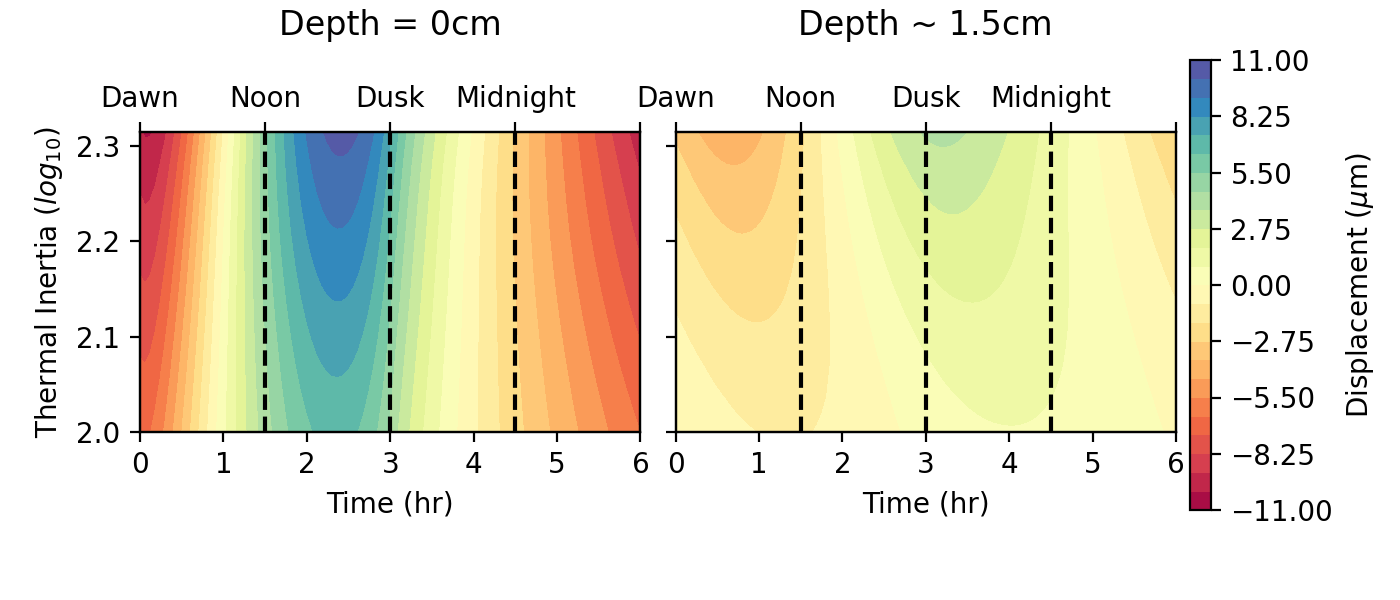}
    \caption{Shown is the variation in vertical displacement $\delta_z$ at two different depths as a function of thermal inertia $\Gamma$ and time throughout the 6-hour rotational period. The panels show integrated displacement $\delta_z$ at two depths, the surface at $z = 0$ cm (on the left) and  $z = 1.5$ cm below the surface (on the right). The x-axis shows time in hours with 0 corresponding to dawn. The y-axis shows the $\log_{10}$ of thermal inertia $\Gamma$ in units of J m$^{-2}$ s$^{-1/2}$ K$^{-1}$. The colorbar shows vertical displacement in microns $\mu $m. This figure was computed using a thermal expansion coefficient of $\alpha_L = 10^{-5}$ K$^{-1}$. Again notice that contours of constant displacement are not vertical and this implies that grains of different thermal inertia in proximity would move with respect to one another.}
    \label{fig:time}
\end{figure}

For a range of thermal conductivities and depths at eight different rotational phases, we show in Figure \ref{fig:dz} the vertical displacement  $\delta_z$, computed from Equation \ref{eqn:dz} assuming an expansion coefficient of $\alpha_L=10^{-5}$ K$^{-1}$. This figure is generated using the temperature profiles shown in Figure \ref{fig:k_T}. In Figure \ref{fig:dz}, the x-axes are thermal inertia, the y-axis is depth and each panel shows a different orbital phase. In Figure \ref{fig:time} the x-axis is the orbital phase in hours (rather than at selected snapshots), the y-axis is thermal inertia, and each panel shows a different depth. In both Figures \ref{fig:dz} and \ref{fig:time} the colorbar shows the displacement $\delta_z$ in microns. Figures \ref{fig:dz} and \ref{fig:time} show that two columns of material with different thermal conductivity would have different functions $\delta_z(z)$ (at a given rotational phase), thus would move up and down with respect to one another. At dawn, the material with the higher thermal conductivity would be lower near the surface than that with lower conductivity, but the opposite trend is seen before dusk.  The two columns would flex periodically with respect to one another.

\begin{figure}
\centering
	\includegraphics[height=45mm]{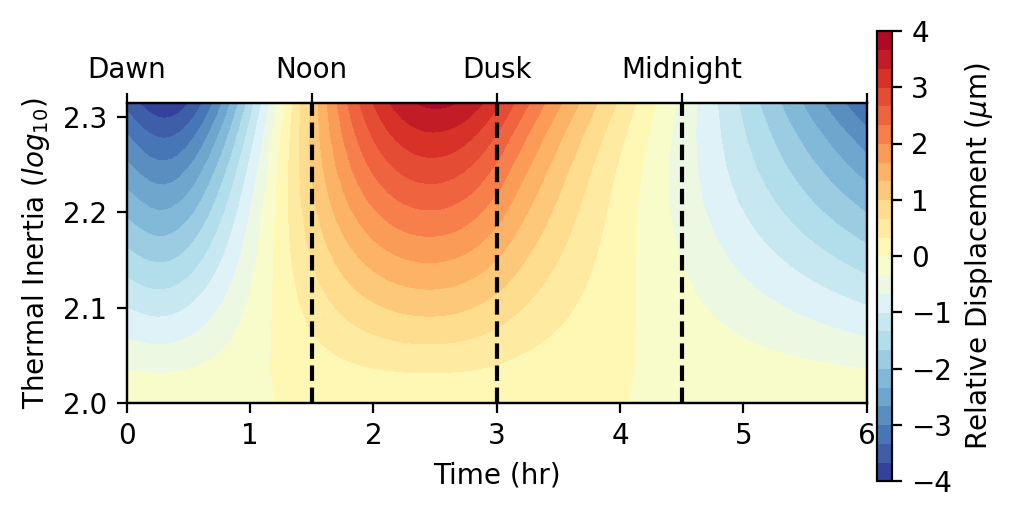}
    \caption{Shown is the variation in relative vertical displacement $\Delta_z$ at the surface as a function of thermal inertia $\Gamma$ and as a function of time. The relative displacement is the difference between that measured for the thermal inertia on the y-axis and that computed for the fiducial model. The x-axis shows time during the 6-hour rotational period. The y-axis shows the $\log_{10}$ of thermal inertia $\Gamma$ in units of J m$^{-2}$ s$^{-1/2}$ K$^{-1}$. The panels show the relative displacement at the surface. The colorbar shows vertical relative displacement at the surface in microns. This figure was computed using a thermal expansion coefficient of $\alpha_L = 10^{-5}$ K$^{-1}$. Again notice that contours of constant displacement are not vertical. This implies that if there are variations in thermal conductivity as a function of depth, grains would move with respect to one another due to variations in thermal expansion.
    \label{fig:reldis}}
\end{figure}

We characterize the extent that particles move with respect to one another with a relative displacement function $\Delta_z(z,\Delta_k)$ that depends on their relative conductivity $\Delta_k$. In Figure \ref{fig:reldis} we plot the relative displacement $\Delta_z$ at the surface $z=0$ as a function of thermal inertia where we use the fiducial model as the reference point. 

Based on the scale given in the colorbars in Figure \ref{fig:dz},  \ref{fig:time}, and \ref{fig:reldis} we find that the size of the relative displacement per cycle is of order a few microns for the maximized variations in local thermal conductivity that are expected for a heterogeneous surface such as asteroid Bennu.

\subsection{Displacements from a different conductivity surface layer}
\begin{figure}
\centering
	\includegraphics[height=38mm]{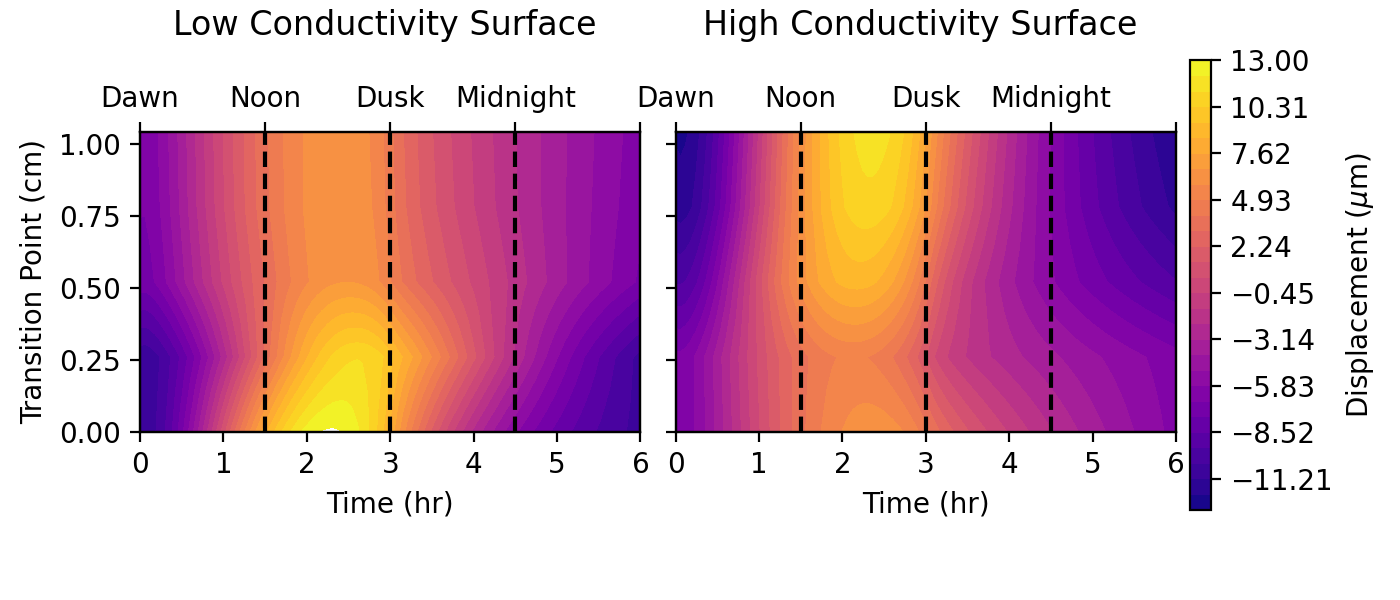}
    \caption{Shown is the effect of having a surface layer of a different thermal conductivity material. The x-axis shows time during the 6-hour rotational period. The y-axis shows the depth of the layer where the conductivity changes in centimeters. The colorbar shows vertical displacement at the surface in microns. The panels show the displacement at the surface for two cases, on the left is shown a low-conductivity material on top of a high-conductivity material, while on the right a high-conductivity material is on top of a low-conductivity material. This figure was computed using a thermal expansion coefficient of $\alpha_L = 10^{-5}$ K$^{-1}$. Again notice that contours of constant displacement are not vertical. This implies that if there are variations in thermal conductivity as a function of depth, grains would move with respect to one another due to variations in thermal expansion.    
    \label{fig:transition}}
\end{figure}

In the previous section, we discussed the simple case where each vertical column consists of a single material but columns in proximity to one another can differ in their thermal properties giving relative motions. This case is shown in Figure \ref{fig:kt_picture} a) (on the left). We now consider the case where conductivity can vary as a function of depth, as shown in Figure \ref{fig:kt_picture} b) (on the right). In this case, at a particular depth, there is a transition to a different conductivity material. Figure \ref{fig:transition} shows the effect this has on the displacement $\delta_z$ at the surface and how this is affected by the depth where the transition between conductivities occurs. The left panel in Figure \ref{fig:transition} has a low conductivity surface, while the right panel has a high conductivity surface. The low conductivity material has thermal conductivity $k_T=10^{-2}$ W m$^{-1}$ K$^{-1}$, consistent with our fiducial model as shown in Table \ref{tab:fiducial}. The high conductivity material has thermal conductivity $k_T=5\times10^{-2}$ W m$^{-1}$ K$^{-1}$, corresponding to the maximum thermal conductivity used in Figures \ref{fig:k_T}, \ref{fig:dz}, and \ref{fig:time} and is otherwise consistent with our fiducial model. Two columns with differences in the thermal conductivity of their layers would also have different functions $\delta_z(z)$ (at a given rotational phase) and would move up and down with respect to one another, similar to what is seen with columns without layered materials.

\subsection{Displacements caused by shadows}
\begin{figure}
\centering
	\includegraphics[height=50mm]{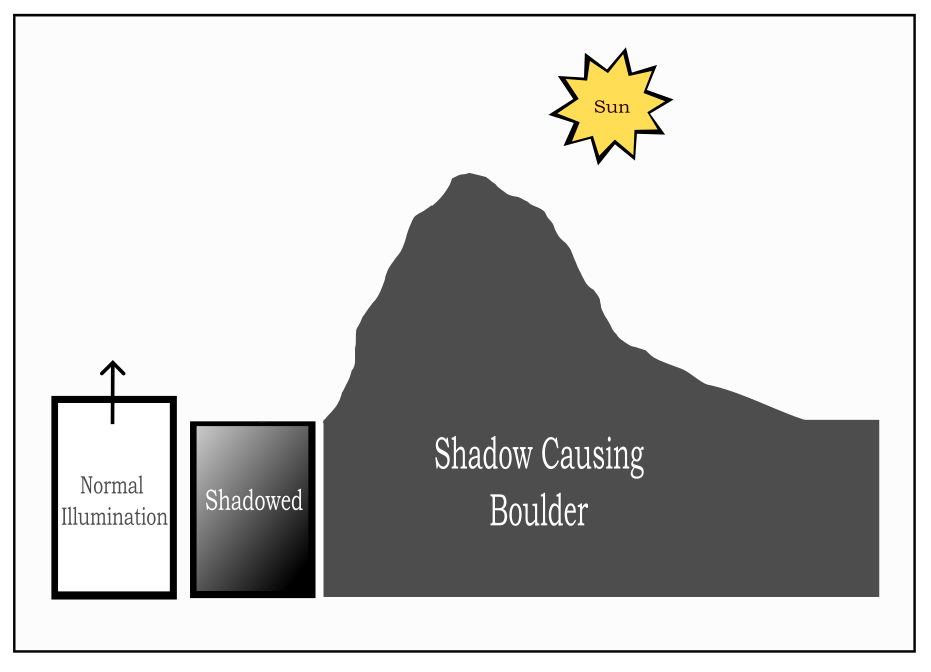}
    \caption{Shadows can cause a similar effect to the variations in conductivity shown in Figure \ref{fig:kt_picture}. Shown are rocks exposed to differences in illumination due to shadows caused by a boulder.  This causes variations in thermal expansion.
    \label{fig:illum_picture}}
\end{figure}

\begin{figure}
\centering
	\includegraphics[height=50mm]{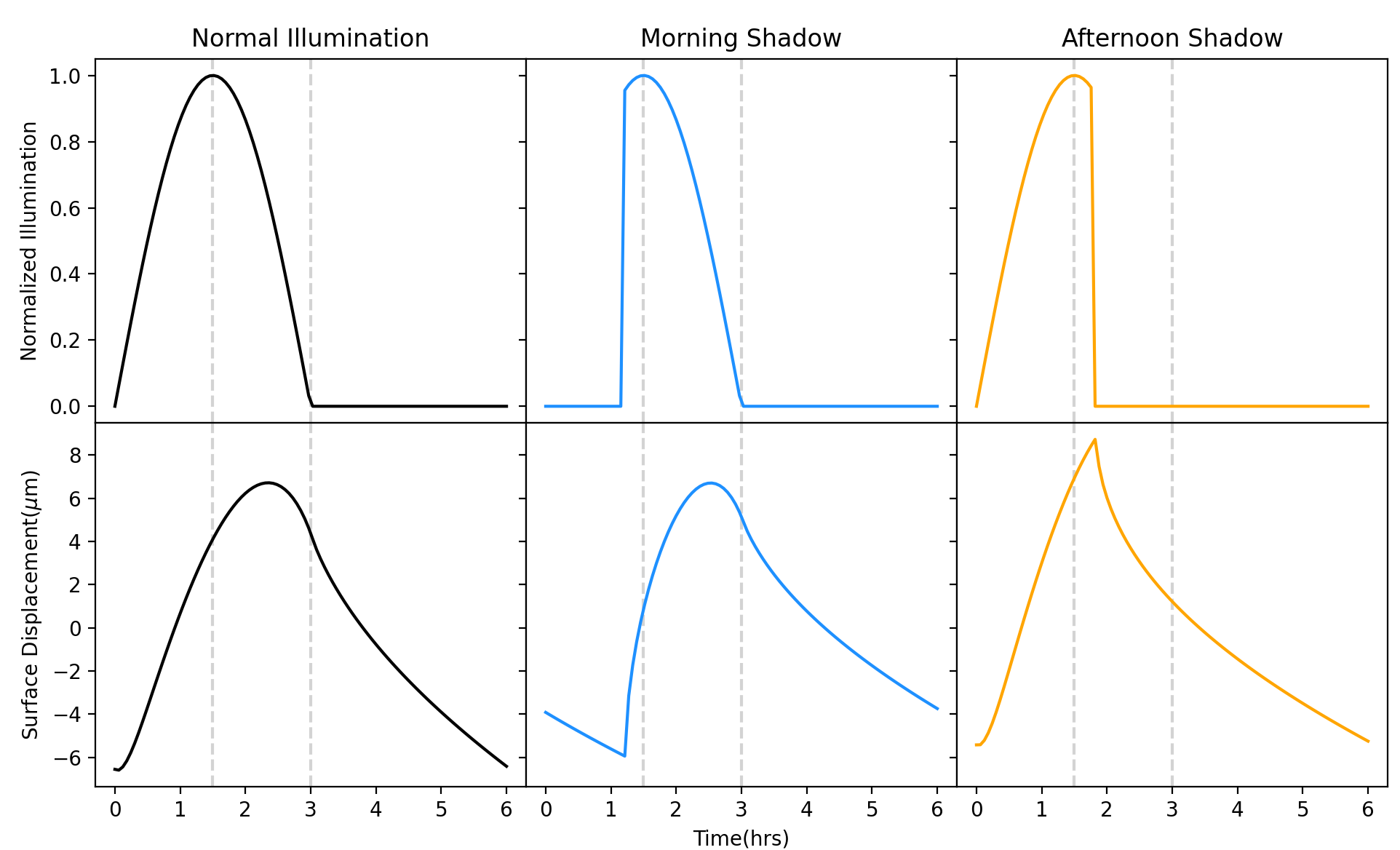}
    \caption{Shown is the effect of daytime shadows; the top row shows the normalized illumination function $S(t) \Psi(t)$ versus time.  The bottom panels show vertical displacement at the surface as a function of time. Left to right is normal illumination used in previous plots, a morning shadow that starts at dawn and ends at 1.2 hours and an afternoon shadow that starts at 1.8 hours and ends at dusk. The x-axis is the same for all 6 plots and shows time in hours throughout a 6-hour spin rotational period. Again this figure was computed using a thermal expansion coefficient of $\alpha_L = 10^{-5}$ K$^{-1}$.
    \label{fig:shadow}}
\end{figure}

In addition to depending on thermal conductivity, the thermal expansion and contraction of a substrate also depend on illumination, as we see throughout the rotational period. Figure \ref{fig:illum_picture} shows how shadows can cause a difference in illumination between nearby regions. Because of the heterogeneity of grain size, larger rocks and boulders could cast shadows on granular media. To simulate shadows, we create more complicated illumination functions than the sinusoidal one used for the fiducial model, given in equation \ref{eqn:illum}. We can shift the shadow throughout the daylight hours to show how shadows at different points in the day affect both temperature and integrated vertical displacement. The illumination function  $S(t)$ is either zero or 1 but we can vary the time of transition. 
The top panels of Figure \ref{fig:shadow} show the difference between the normal illumination function used in the fiducial model and the illumination functions from morning and afternoon shadows. 
The bottom panels show the resulting displacement for each case. A shadow present in the morning would delay the heating of the surface, while a shadow present in the afternoon hastens nighttime cooling. Two columns that receive different amounts of illumination throughout their 6-hour period would have different functions $\delta_z(z)$ (at a given rotational phase) and would move up and down with respect to one another. The size of the relative displacement caused by shadows is similar to what is seen when we varied the conductivity in Figure \ref{fig:time}. 

\section{Discussion}
\label{chap:chapter-3}

\subsection{Diffusive behavior caused by thermal cycling}
Molecular dynamics simulations of thermal cycling in granular systems have suggested that particle motions are diffusive \citep{Percier_2013}.
Assuming that diffusive behavior caused by thermal cycling takes place on asteroids, we estimate the maximum distance that a particle can travel in a given amount of time. 

Because the magnitude of the temperature change $\Delta T$ in each thermal cycle depends on the distance to the Sun, we expect $\Delta T= T_{noon} - T_{midnight} \propto r_H^{- \frac{1}{2}} $ (using equation \ref{eqn:noon}). Since our surface displacement depends on the change in temperature that causes the thermal expansion, this gives surface displacement due to thermal expansion $\delta_z \propto r_H^{- \frac{1}{2}}$.
As mentioned previously, the skin depth
characterizes the depth of material that undergoes temperature variations due to thermal cycling. Consequently, we expect the height variation 
$\delta_z \propto l_{2\pi} \propto \sqrt{ \frac{P_{rot} k_T }{\rho C_P}}$.
If the relative displacement $\Delta_z$ between two grains is caused by variations in thermal conductivity, then we expect 
$\Delta_z \propto \sqrt{\frac{\Delta_K}{k_T}}$ where $\Delta_K$ characterizes the differences in thermal conductivity. Putting together these factors, we expect 
\begin{align}
\Delta_z \sim & 8\, \mu {\rm m} 
\left(\frac{P_{rot}}{6 \ {\rm hr}}  \right)^\frac{1}{2}
\left( \frac{\rho}{2000\ {\rm kg\ m}^{-3}} \right)^{-\frac{1}{2}}  \times 
\nonumber \\ & \ \ \ 
\left( \frac{C_P}{500\ {\rm  J\ kg}^{-1} {\rm K}^{-1}} \right)^{-\frac{1}{2}} 
\left( \frac{k_T}{0.01\ {\rm W\ m}^{-1} {\rm K}^{-1} } \right)^\frac{1}{2}  \times 
\nonumber \\ & \ \ \ \ 
\left(\frac{r_H}{1 \ {\rm AU} } \right)^{-\frac{1}{2}} 
\left( \frac{\alpha_L}{ 10^{-5} {\rm K}^{-1} } \right)
\left(\frac{\Delta_K}{5} \right)^\frac{1}{2} 
\end{align}
where the $8 \mu m$ value is approximately taken from Figure \ref{fig:dz} at $z=0$ (the surface) for the largest difference in displacement (at dawn or dusk) between the lowest and highest conductivity materials. 

Suppose the relative displacement $\Delta_z$ characterizes the size of motions in a random walk, as suggested by the 2D molecular dynamics simulations of thermal cycling in granular systems by \citet{Percier_2013}. Following \citet{Percier_2013}, we model ratcheting with a diffusion coefficient $D$ that depends on the size of the relative displacement and the cycling period 
\begin{equation}
D = \frac{\Delta_z^2}{P_{rot}} . \label{eqn:D}
\end{equation}
Assuming the thermal cycling gives rise to a diffusive behavior, the distance that a grain travels during a time $t$ is 
\begin{equation}
    d \sim \sqrt{D t}, \label{eqn:d_Dt}
\end{equation}
giving
\begin{equation}
d \sim 0.4\ {\rm m} \left( \frac{\Delta_z}{10 \mu {\rm m} } \right)
\left( \frac{P_{rot} }{\rm 6\, hr} \right)^{-\frac{1}{2}}
\left( \frac{t}{\rm Myr} \right)^\frac{1}{2}.
\end{equation}
In the most optimal conditions, in a million years, diffusive behavior could give transport for particles in the skin depth layer significantly further than the skin depth itself. 

Continuing to follow \citet{Percier_2013}, diffusive behavior depends on the variation in grain size due to thermal cycling  which is approximately $\alpha_L \Delta T$.  \citet{Percier_2013} compare this quantity to the ratio of the gravitational force on a grain to the product of elastic and frictional forces. Because the gravitational acceleration is low on an asteroid, the cycling amplitude (in terms of temperature) vastly exceeds the ratio of weight $mg$ to an elastic or friction force throughout the thermal skin depth where $m$ is a grain mass and $g$ the gravitational acceleration. We can compare $\alpha_L \Delta T$ to the ratio of hydrostatic pressure at the depth of the thermal skin depth $\rho g l_{2\pi}$ and the elastic modulus of the medium; 
\begin{align}
 \frac{\rho g l_{2\pi}  }{E} 
 & = 10^{-10} \left(\frac{\rho}{2000~ {\rm kg\ m}^{-3}} \right)
 \left(\frac{g/g_\oplus}{10^{-4} } \right) 
 \left(\frac{l_{2\pi}}{5 \ {\rm cm}} \right)  \times \nonumber \\
 & \ \ \ \ 
 \left(\frac{E}{ 10^8 {\rm Pa}} \right)^{-1} .
\end{align}
This ratio is really small, implying that thermal cycling  would likely give stresses that allow motion throughout the thermal skin depth. If the diffusive behavior seen in the simulations by \citet{Percier_2013} applies to an asteroid, we would expect the entire skin depth to flow due to thermal cycling, rather than exhibit stick-slip behavior. These estimates suggest that in the right conditions, pebbles can migrate diffusively a distance of order a few centimeters over a million years due to the expansion and contraction that takes place during thermal cycling.

%%%
This expression should also include a factor that is dependent on the difference in thermal inertia or thermal conductivity of nearby rocks. Here we put $\Delta_K$ as the ratio of high to low thermal conductivity among the different rocks. Note that in equation \ref{eqn:D} the spin period cancels out due to the dependence of $\Delta_z$ on $P_{rot}^\frac{1}{2}$ giving a diffusion coefficient that is independent of the spin period.  This implies that creep due to thermal cycling would not preferentially occur on fast rotators.

\subsection{Ratcheting}
By applying vertical stress cycles in a 2-dimensional simulation of polygons, \citet{Alonso_Marroqin_2004} show that stress cycling in a granular system can cause a ratcheting effect. The micromechanical origin of this effect is due to fluctuations in the complex network of contacts. Even small variations in stress can allow tangential contacts to reach a sliding condition giving irreversible motions. Because of the load-unload asymmetry of the contact forces, a net accumulation of plastic deformation can be observed in each cycle. This leads to a constant increase in plastic strain.

Another possibility is that ratcheting of grains not only gives diffusive behavior but allows downhill creep (e.g., \cite{Deshpande_2021}). If the distance moved by a grain via ratcheting is linearly dependent on time, then the maximum distance a grain could travel would be as large as,
\begin{align}
d &\sim  \frac{\Delta_z}{P_{rot}} t  \nonumber \\
& = 14.6\ {\rm km} \left( \frac{\Delta_z}{10\mu {\rm m}} \right) 
\left(\frac{P_{rot}}{6\ {\rm hr}} \right)^{-1}
\left( \frac{t}{\rm Myr} \right) .
\end{align}
This linear dependence on time contrasts with that estimated diffusively in equation 
\ref{eqn:d_Dt} where no direction is preferred.  If transport occurs preferentially in a particular direction, then it is possible for particles in the right conditions to travel large distances across the surface. In contrast, diffusive behavior would arise because of stochastic rearrangements in the force and contact networks rather than ratcheting in a preferred direction. 

Over long periods of time, material within the thermal skin depth would creep or flow downhill. The surface of asteroids contains regions with downhill slopes. Creep associated with thermal cycling gives another possible mechanism that could account for the formation of terraces on Bennu \citep{Barnouin_2022b} that are held up by larger boulders or collections of fine-grain materials in depressions. % or along the equatorial region.  {\bf  citations?  Bennu has finer materials on equator}
\subsection{Compaction}
Experiments and simulations show that thermal cycling leads to compaction with packing fraction increasing by a few percent over tens of cycles \citep{Chen_2006, Chen_2009, Vargas_2007}. Past a characteristic number $n_c \sim 100$ of cycles, the packing fraction continues to increase but more slowly, and logarithmically,  as typical of a relaxation mechanism \citep{Chen_2006, Chen_2009, Vargas_2007, Divoux_2008, Divoux_2010, Blanc_2013}. Experiments have been carried out for a few thousand cycles, suggesting that the packing fraction would continue to slowly increase with even higher numbers of cycles \citep{Divoux_2008, Divoux_2010, Blanc_2013}.

\citet{Chen_2006} and \citet{Chen_2009} carried out thermal cycling experiments on spherical glass grains in cylindrical glass containers. The initial packing fraction for both experiments was approximately 58.9$\%$. A full thermal cycle was roughly 20 hours with temperature differences of roughly 100 K and 40 K. Dozens of thermal cycles were performed and both cycling temperatures behaved logarithmically, but the packing fraction for the 100 K cycling temperature initially increased more rapidly but fell off sooner. A similar simulated result was found by \citet{Vargas_2007} for similar cycling temperatures and experimentally by Blanc \citet{Blanc_2013} and \citet{Divoux_2008} for small cycling temperatures. In \citet{Chen_2006} and \citet{Chen_2009} the packing fraction verse cycles was fit to a double relaxation function, $y=y_0-A_1 e^{-x/t_1}-A_2 e^{-x/t_2}$.  Extrapolating out to 100 cycles the packing fraction had increased by $\sim$5$\%$ and subsequent cycles increased the packing fraction by less than a percent. For smaller cycling temperatures more cycles impacted the packing fraction but resulted in the same amount of compaction after enough time. Because the large cycling temperatures on asteroids are large if sediment was loosely packed enough for compaction to occur, it would happen very rapidly compared to the lifetime of asteroids. Compaction is also aided by gravity, so on asteroid's surfaces the change in packing fraction might not be as significant.
\subsection{Penetration}
With the Brazil Nut Effect, vibrations allow size segregation, moving larger grains to the surface above smaller grains. \citet{Chen_2009} studied the displacement from thermal cycling of an "intruder" grain introduced into a container of uniform grains. The uniform grains were made of polystyrene while the intruders were made of brass and Teflon with varying diameters. \citet{Chen_2009} found that unlike with the Brazil Nut Effect, these intruders sank further below the surface. The higher the cycling temperature, the deeper the intruding object would sink. They also found that the depth penetrated depended both on the intruder's size and density. Intruders with lower density and smaller diameters than the surrounding grains were not able to penetrate as deeply. This does not match what is expected from the Brazil Nut Effect where larger grains rise, smaller grains sink, and there is little or no dependence on grain density \citep{rosato1987}.  

\citet{Blanc_2013} conducted a similar experiment to \citet{Chen_2009} but using containers with mixed grain diameters and did not track individual grains. Both studies, as mentioned in the previous section, found that all of the grains compactified. This suggests that the penetration of certain grains and the potential size segregation is part of the processes of compaction of the granular media. Regardless, if denser grains tend to sink, this could lead to more porous grains at the surface, possibly explaining unexpectedly low thermal inertia values \citep{Rozitis_2020}, especially at the equator where there would be the highest day-night variation in temperature. 
\subsection{Caveats}
It is important to note that we used a  1-dimensional thermal conductivity model to estimate the size of relative displacements, so we neglected the role of thermal conductivity in 2 dimensions. Anywhere there is contact between adjacent rocks, heat would travel between them. Heat could also be transferred via radiation between them \citep{Persson2022}. However, because the heat source is at the surface the primary heat flow would be perpendicular to the surface in the z-direction. We also neglect the dependence of thermal conductivity on temperature in our model (e.g., \cite{Vasavada_2012}).  These effects are unlikely to affect the estimate of the relative displacement size to order of magnitude but could be investigated in future work.

There are additional processes that reshape the surface of asteroids. Mass flows associated with spinup due to the YORP effect, tidal perturbations from planetary encounters, thermal fracturing, and impacts can also displace surface materials \citep{DeMeo2023}. Future work can compare the role of transport associated with thermal cycling to these other processes. 

As mentioned previously, laboratory experiments of thermal cycling in granular materials have been conducted for industrial applications \citep{Divoux_2010, Chen_2009, Blanc_2013}. The conditions for these studies are not similar to conditions on asteroid surfaces. In these experiments, they attempt to heat the grains evenly \citep{Divoux_2010, Chen_2009, Blanc_2013}. However, on asteroid surfaces, the grains are not heated as evenly, because heat propagates from the surface due to illumination. This causes differences in the expansion of the grain with depth. Most experiments and simulations have only been carried out for at most 1000 cycles \citep{Divoux_2010, Chen_2009, Blanc_2013} and it is difficult to extrapolate to millions of thermal cycles. In many of these experiments, the grains are smooth spheres \citep{Divoux_2010, Chen_2009, Blanc_2013}. The grains on asteroids are varied and have more jagged shapes than spheres or disks. They also have more varied grain sizes than those used in these experiments. This might explain why grains ratcheting upward were not seen in the experiments by \citet{Chen_2009}. 

Asteroid surfaces are low-g and airless  environments. While experiments of thermal cycling granular media have not been conducted in microgravity conditions, parabolic flights have shown that gravity has an effect on granular flow. \citet{Murdoch2013} found that while the primary flow in their system was unaffected by gravity, a secondary convective flow was nonexistent in the absence of the Earth's strong gravity. Because \citet{Chen_2009} found penetration and size segregation of grains depends on density, reduced gravity could affect this effect as well. Further simulations and experiments could help us to understand the effect that the microgravity and airless environment of asteroid surfaces would have on thermal cycling granular material.

%%%%%%%%%%%%%%%%%%%%%%%%%%%%%%%%
\section{Conclusions}
\label{chap:chapter-4}
In this manuscript, we have explored the relative motions of grains on an asteroid's surface due to thermal cycling. Recent observations show that material on asteroid surfaces has a wide range of thermophysical properties. Based on the variations in thermal inertia seen on Ryugu and Bennu, pebbles in proximity to one another could have thermal conductivities that differ by a factor of 5. This implies that grains in proximity experience different levels of temperature extremes which would give relative motions between them. With a 1-dimensional thermophysical model, we estimate the relative difference in surface height between two columns of materials with different conductivity.  We find that the relative height varies by about a few microns during an asteroid's spin rotation period. We also looked at layers of materials with different conductivity and the effect of grains receiving differences in illumination during the day due to shadows. These variations produced a similar size relative displacement. 

If grains are in contact, the variations in relative displacement would cause them to them ratchet against each other. Even though this relative displacement of a few microns is small, the large number of thermal cycles (millions to billions) taking place on an asteroid might make this process important within the thermal skin depth, which is the top few centimeters of a rubble surface. If the ratcheting behavior is diffusive, we find that pebbles in the right conditions can move up to a few centimeters in a million years. This implies that the skin depth of an asteroid with a granular surface could evolve slowly due to relative motions induced by thermal cycling. Thermal cycling experiments also tend to show creep in the downhill direction (e.g., \cite{Deshpande_2021}), which could allow surface particles to travel even further than a few centimeters. 

Laboratory experiments in granular media find that thermal cycling can cause compaction \citep{Divoux_2010}, larger and more dense grains to sink \citep{Chen_2009}, and  downhill creep \citep{Deshpande_2021}. These effects are enhanced by gravity and might be diminished on asteroid surfaces because of their lower gravity. However, we estimate that the ratio between stress due to thermal cycling and hydrostatic pressure is higher on an asteroid than on the Earth. This suggests that diffusive transport due to thermal cycling could be more effective in low g environments than on the Earth.

If cumulative, over the lifetime of an asteroid, transport associated with thermal cycling could have an impact on an asteroid's surface. Future work could determine whether this process is important compared to other resurfacing processes, such as spinup from the YORP effect, tidal perturbations during planetary encounters, thermal fracturing, and impacts \citep{DeMeo2023}. Current experiments and simulations (e.g., \cite{Alonso_Marroqin_2004, Chen_2009, Blanc_2013, Divoux_2010}) give us an idea of how a thermal cycled granular medium would behave on an asteroid surface. However, more work is needed to determine the effect of microgravity and airless environments on thermal cycling induced granular flow.

\section{Acknowledgements}
\noindent
This material is based upon work supported in part by Craig McMurtry and by the University of Arizona through the Near-Earth Object Surveyor Mission, Phase B under NASA Grant 80MSFC20C0045, by Cornell University through New York Space Grant 80NSSC20M0096 and to A. C.  Quillen through NASA grant 80NSSC21K0143.

\bibliographystyle{elsarticle-harv}
\bibliography{refs_atten.bib}

\appendix
\section{1-D Thermophysical Model}

\label{sec:append}
We follow the 1-dimensional thermal conductivity model and the procedure for computing it that is summarized in section 2.2 by \citet{Rozitis_2011}. The temperature is a function of depth and time $T(z,t)$ where $z>0$ is the depth below the surface. The surface lies at $z=0$. 
Conservation of energy on the surface gives the surface boundary condition:
\begin{align}
(1 - A_B) ( [1 - S(t)] \Psi(t) F_{SUN} + F_{SCAT} ) &  \nonumber \\
\qquad 
\qquad + ( 1 - A_{TH}) F_{RAD} 
+ k_T \frac{dT}{dz}_{z=0} & 
\nonumber \\
\qquad  - \varepsilon \sigma_{SB} T^4_{z=0} & = 0. \label{eqn:surface}
\end{align}
Here  $\varepsilon$ is the emissivity, $\sigma_{SB}$ is the Stefan-Boltzmann constant, $A_B$ is the Bond albedo, $S(t)$ indicates how the surface is shadowed at time $t$,  and $k_T$ is the thermal conductivity. The shadow function is $S(t)= 1$ if the surface is in shadow, and is 0 otherwise. The energy flux 
$F_{SUN} =  F_\odot/r_H^2$ is the integrated (over wavelength) solar flux at the distance of the object from the Sun,
with solar constant $F_\odot = 1367$ W m$^{-2}$ and  $r_H$ equal to the heliocentric distance of the body in AU. The fluxes 
$F_{SCAT}$ and $F_{RAD}$ are the total scattered and thermal radiated fluxes incident 
on the surface region.
%(meaning they are hitting the facet, and not being emitted by the facet).
The albedo of the surface at infrared wavelengths is $A_{TH}$. The function 
$\Psi(t)$ returns the cosine of the Sun's illumination angle at a time $t$, which depends on the facet and rotation pole orientations, and it changes periodically as the planetary body rotates.
For our model we set $F_{SCAT} = 0$, $F_{RAD} = 0$,  and $r_H=1 AU$.

Below the surface,  heat conduction is described in 1-dimension. In the absence of internal heat sources, the heat equation is; 
\begin{equation}
\frac{ \partial T}{\partial t} = \frac{k_T}{\rho C_P}  \frac{\partial^2 T}{\partial z^2}  
= \kappa  \frac{ \partial^2 T}{\partial z^2}, \label{eqn:heat}
 \end{equation}
where $\rho$ is density, $C_P$ is specific heat and $\kappa =  \frac{k_T}{\rho C_P}$ is the thermal diffusivity. 
We discretize the temperature $T(z,t)$ in space and time. At each grid position and time, the temperature is $T_i^n$, with $i$ index specifying  the spatial position and $n$ referring to the time. The grid spacing is even in both space and time with $dz$ giving the distance between depth positions and $dt$ the duration of each time step. 

The heat diffusion equation (equation \ref{eqn:heat}) is integrated with an explicit scheme;
\begin{equation}
T_j^{n+1} = T_j^n + \frac{\kappa dt}{dz^2} ( T_{j+1}^n + T_{j-1}^n - 2 T_j^n),  \label{eqn:scheme}
\end{equation}
which requires $dt <  dz^2/\kappa$ for numerical stability.
Equation \ref{eqn:scheme} is used to update the temperature at all points in the
grid, except for the bottom-most and the top-most (or surface) grid points. 
With $N$ points in the spatial grid, our grid index ranges from $i=0$ to $i=N-1$. 

At the bottom of the spatial grid we assume a zero flux boundary condition
\begin{equation}
\frac{\partial T}{\partial z} \Big|_{bottom}  = 0.
\end{equation}
The lower boundary condition is
equivalent to 
\begin{equation}
T_{N-1}^{n+1} = T_{N-2}^{n+1}. \label{eqn:bottom}
\end{equation}
After equation \ref{eqn:scheme} is used to update the temperature at all grid points except the bottom and top, Equation \ref{eqn:bottom} is used to update the bottom grid point temperature.

Next is to find a way to update $T_0^{n+1}$, the temperature at the surface
using equation \ref{eqn:surface}.
We define a function for the surface temperature 
\begin{align}
f(T) & = (1-A_B)(1- S(t)) \Psi(t) \frac{F_\odot}{r_H^2}  + \frac{k_T}{dz} (T_1^{n+1} - T) \nonumber \\
& \qquad - \epsilon \sigma_{SB} T^4 
\label{eqn:f}
\end{align}
where $T_1^{n+1}$ is the temperature at the grid point just below the surface grid point after we
have updated the temperature array with equation \ref{eqn:scheme}. 
The function $f(T) =0$ for a surface temperature that is consistent with the boundary
condition in equation \ref{eqn:surface} (but neglecting scattering, shadows and radiative heat).
The Newton-Raphson technique is used to iteratively find the surface temperature.
Starting with $T = T_0^n$ we iteratively perform the operation 
\begin{equation}
T_{R+1} = T_R - \frac{f(T_R)}{f'(T_R)} .
\end{equation} 
After about 10 iterations, $f(T)=0$ is satisfied at high precision giving us a value for $T_0^{n+1}$ and completing a full update of the temperature array. 

The illumination function we use is
\begin{align} 
\Psi( \tau) &= \Bigg\{  \begin{array}{ll} \sin (2 \pi \tau)  \\ 0 \\ \end{array} 
 \begin{array}{ll} {\rm for} \\ {\rm for} \\ \end{array}  
 \begin{array}{ll} 0 \le \tau  <  0.5 \\
 0.5 \le \tau   <  1 \\ \end{array}  \label{eqn:illum}
 \end{align}
with 
\begin{align} 
 \tau &= \frac{t}{P_{rot}} \mod (1) 
\end{align}
giving the orbital phase, and $P_{rot}$ equal to the spin rotation period. 
Here noon corresponds to $\tau = 1/4$ and dawn is at $\tau=0$.  We only consider equatorial illumination for a spinning object with obliquity of zero and ignore the poles. 

It is convenient to define a thermal skin depth at which the phase lag of the internal temperature
variation is about $2\pi$ 
\begin{equation}
l_{2 \pi}  = \sqrt{\frac{4 \pi P_{rot} k_T}{\rho C_P} }. \label{eqn:skindepth} 
\end{equation}
We use a grid that has maximum depth at least twice the skin depth. 
%where $P_{rot}$ is the spin rotation period. 

In the absence of shadows, scattering, absorption of radiated thermal heat from nearby surfaces, and heat flux from below,
equation \ref{eqn:surface}
gives an estimate for the temperature at noon.
\begin{equation}
T_{noon}  \sim \left( \frac{(1.0 - A_B) F_{SUN} }{ \varepsilon \sigma_{SB} } \right)^\frac{1}{4}. \label{eqn:noon}
\end{equation}
We initialize the temperature array with $T_{noon}$ at the surface. We set the temperature to be $T = T_{noon} 2^{-\frac{1}{4}}$ at the bottom of the grid and linearly interpolate between the surface and base. Starting with this initial condition for $T$ and at $\tau=0$, corresponding to dawn, we integrate the system, by consecutively updating the temperature array at each time step for a total of 200 spin rotation  periods. Note that the surface boundary condition is time-dependent due to the illumination function. After this integration time, transients in the system have decayed and the temperature gradient at the base is constant. 
After transients have decayed we can compute $T(z,\tau)$ at any desired orbital phase $\tau$ by integrating for an additional length of time equal to $\tau P_{rot}$.

\end{document}